\documentclass[aps,twocolumn,floatfix,amsmath,amssymb]{revtex4-1}

\usepackage{amsmath,bm,amsfonts}
\usepackage{graphicx}

\newfont{\bg}{cmr10 scaled\magstep4}

\newcommand{\bigzerou}{\smash{\lower1.7ex\hbox{\bg 0}}}

\begin{document}

\title{Efficient O($N$) divide-conquer method with localized natural orbitals}
  
\author{Taisuke Ozaki and Masahiro Fukuda}
\affiliation{
  Institute for Solid State Physics, The University of Tokyo, Kashiwa 277-8581, Japan
}

\author{Gengping Jiang}
\affiliation{
  College of Science, Wuhan University of Science and Technology, Wuhan, 430081, China
}

\date{\today}

\begin{abstract} 
An efficient O($N$) divide-conquer (DC) method based on localized natural orbitals (LNOs) is presented 
for large-scale density functional theories (DFT) calculations of gapped and metallic systems. 
The LNOs are non-iteratively calculated by a low-rank approximation via a local eigendecomposition of a projection operator 
for the occupied space. 
Introducing LNOs to represent the long range region of a truncated cluster reduces the computational cost 
of the DC method while keeping computational accuracy. 
A series of benchmark calculations and high parallel efficiency in a multilevel parallelization clearly demonstrate
that the O($N$) method enables us to perform large-scale simulations for a wide variety of materials 
including metals with sufficient accuracy in accordance with development of massively parallel computers.

\end{abstract}

\maketitle

\section{INTRODUCTION}

First-principles electronic structure calculations based on the density functional theories (DFT) \cite{Hohenberg1964,Kohn1965} 
have been playing a versatile role in a wide variety of materials sciences to deeply understand physical and chemical properties of 
existing materials and even to design novel materials having a desired property before actual experiments \cite{Neese2009,Jones2015,Cole2016,Jain2016}. 
In recent years more complicated materials with secondary structures such as heterointerfaces \cite{Sawada2013,Sawada2017} and 
dislocations \cite{Wakeda2017} have been becoming 
a scope of application by DFT calculations. Since these complicated structures cannot be easily modeled by a small unit cell, 
development of efficient DFT methods in accordance with development of massively parallel computers is crucial to realize 
such large-scale DFT calculations. 
Among efficient DFT methods \cite{Lin2009,Ozaki2010,Goedecker1999,Bowler2012,Aarons2016,Soler2002,Gillan2007,Skylaris2005,Tsuchida2007,Shimojo2008,Mohr2015,VandeVondele2005,VandeVondele2012,Khaliullin2013,Krajewski2005,Inglesfield1981,Ho2008,Wang1995,Zeller1995,Zeller2008,Lin2018,Yang1991,Yang1995,Ozaki2001,Ozaki2006}, 
O($N$) methods whose computational cost scales linearly as a function of the number of atoms 
have enabled us to extend the applicability of DFT to large-scale systems 
\cite{Goedecker1999,Bowler2012,Aarons2016,Soler2002,Gillan2007,Skylaris2005,Tsuchida2007,Shimojo2008,Mohr2015,VandeVondele2005,VandeVondele2012,Khaliullin2013,Krajewski2005,Ho2008,Inglesfield1981,Wang1995,Zeller1995,Zeller2008,Lin2018,Yang1991,Yang1995,Ozaki2001,Ozaki2006}. 
Nevertheless applications of the O($N$) methods to metallic systems have been still limited because of the fundamental 
difficulty of a truncation scheme in real space, which is an idea commonly adopted in most of the O($N$) methods 
\cite{Goedecker1999,Bowler2012}, in realizing the O($N$) methods \cite{Aarons2016} as discussed in the Appendix.
A theoretically proper approach to go beyond the truncation scheme is to take account of the contribution 
from the external region beyond the truncated region via a self energy in Green function formalisms 
\cite{Inglesfield1981,Wang1995,Zeller1995,Zeller2008,Lin2018}. 
Another straightforward approach is to use a relatively large cutoff radius in the truncation scheme 
in order to reach a sufficient accuracy. The latter approach might be suited to the divide-conquer (DC) method \cite{Yang1991,Yang1995}
among the O($N$) methods proposed so far in the following twofold aspects: 
(i) There is a way to reduce the computational cost in the framework of the DC method by lowering 
the dimension of matrices with an introduction of a Krylov subspace \cite{Ozaki2001,Ozaki2006}. 
(ii) The computational time of the DC method can be reduced by a massive parallelization, since the calculations 
in the DC method are performed nearly independently for each atom \cite{Shimojo2005,Duy2014}. 
With the two aspects the improvement of the DC method can be a promising direction to develop an accurate, efficient, 
and robust O($N$) method applicable to not only insulators and semi-conductors, but also metals. 
Along this line, the DC method based on the Krylov subspace has been proven to be an efficient and accurate O($N$) method 
by a wide variety of applications such as dynamics of Li ions in a lithium ion battery \cite{Ohwaki2012,Ohwaki2014,Ohwaki2018} 
and structure optimization of semicoherent heterointerfaces in steel \cite{Sawada2013,Sawada2017}. 
However, there exist drawbacks in the generation of the Krylov subspace \cite{Ozaki2006}. 
Since the Krylov subspace is generated at the first self-consistent field (SCF) step and kept unchanged 
during the subsequent SCF calculation, 
calculated quantities such as the electron density and total energy depend on the initial guess of electron density or 
Hamiltonian matrix elements. If the Krylov subspace is regenerated every SCF step to avoid the dependency on 
the initial guess, the computational efficiency must be largely degraded.
In addition the iterative calculations in the generation of the Krylov subspace tend to suffer from numerical 
round-off error, leading to an uncontrollable behavior in the SCF calculation if the Krylov subspace is regenerated 
every SCF step \cite{Ozaki2006}. 
Therefore, a more robust approach needs to be developed to improve the DC method, which overcomes the drawbacks 
inherent in the DC method based on the Krylov subspace. 

In this paper, we focus on localized natural orbitals (LNOs) calculated by a low-rank approximation
to perform a coarse graining of basis functions, and apply the LNOs to the DC method, which will be referred to as 
the DC-LNO method hereafter, to reduce the dimension of matrices without sacrificing the accuracy. 
The DC-LNO method overcomes the drawbacks in the DC method based on the Krylov subspace, while taking account of reduction of 
the prefactor in the computational cost and a simple algorithm with less communication leading to a high parallel efficiency. 
A series of benchmark calculations clearly demonstrates that the DC-LNO method is an accurate, efficient, and robust O($N$) 
method applicable to not only insulators and semi-conductors, but also metals in step with recent development of massively 
parallel computers.
 
The paper is organized as follows: In the section II we propose a method to generate LNOs, and present the DC-LNO method 
as an extension of the DC method. In the section III the implementations of the method are discussed in detail. 
In the section IV a series of benchmark calculations is presented. In the section V we summarize the 
theory of the DC-LNO method and numerical aspects.

\section{THEORY}

\subsection{General}

 We consider an extension of a divide-conquer (DC) approach \cite{Yang1991,Yang1995} by introducing a coarse graining 
 of basis functions as shown in Fig.~1. For each atom the Kohn-Sham (KS) Hamiltonian and overlap matrices are truncated 
 within a given cutoff radius, and the resultant truncated cluster problem is solved atom by atom, leading to the O($N$) scaling 
 in the computational cost. As expected the error of truncation can be systematically reduced in exchange for 
 the increase of computational cost as the cutoff radius increases.
 The number of atoms in a truncated cluster exceeds 300 atoms in many cases in order to attain a sufficient accuracy 
 as discussed later on. To reduce the computational cost we introduce a coarse graining of basis functions that 
 the original basis functions, pseudo-atomic orbitals (PAOs) in our case \cite{Ozaki2003,Ozaki2004}, are replaced 
 by localized natural orbitals (LNOs) 
 in the long range (yellow) region to represent the Hamiltonian and overlap matrices, while the PAO functions remain unchanged 
 in the short range (orange) region.  
 In the following subsections a method of generating LNOs and a DC method with LNOs are discussed in detail. 

\begin{figure}[t]
    \centering
    \includegraphics[width=8.0cm]{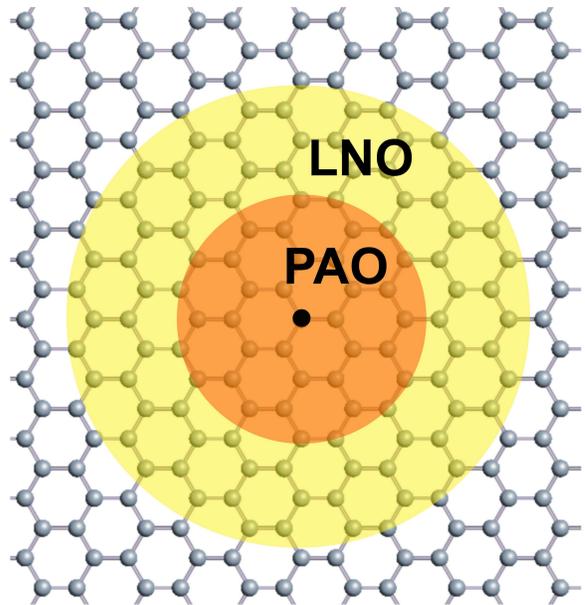}
    \caption{
     Truncation of a system in the DC method with LNOs. The short range (orange) and 
     long range (yellow) regions are represented by PAOs and LNOs, respectively.  
    }
\end{figure}

\subsection{Generation of LNOs}

 We present a method of generating LNOs based on a low-rank approximation via a local eigendecomposition 
 of a projection operator. The method might be applicable to any local basis functions, and not limited 
 to the application to the DC method we discuss in the paper. 
 Even in the conventional O($N^3$) calculations, the LNOs can be easily obtained by a non-iterative calculation
 using the density matrix and overlap matrix elements. Since a smaller number of LNOs well reproduce dispersion 
 of occupied bands, they can be an alternative basis set for a compact representation for the Hamiltonian and 
 overlap matrices. Thus, we present the formulation in a general form as below. 
 
 Under the Born-von Karman boundary condition 
 we expand a Kohn-Sham (KS) orbital $\phi_{{\bf k}\mu}$, 
 indexed with a k-vector ${\bf k}$ in the first Brillouin zone and band index $\mu$, using PAOs $\chi$ \cite{Ozaki2001,Ozaki2006}, 
 being a real function, as 
 \begin{eqnarray}
    \vert \phi_{{\bf k}\mu}\rangle
     = 
    \frac{1}{\sqrt{N_{\rm BC}}}
    \sum_{{\bf R}}
    {\rm e}^{{\rm i}{\bf k}\cdot {\bf R}}
    \sum_{i\alpha} 
    c_{ {\bf k} \mu,i\alpha} 
    \vert \chi_{{\bf R} i\alpha} \rangle,
 \end{eqnarray}
 where ${\bf R}$, $N_{\rm BC}$, and $c$ are a lattice vector, the number of cells in the boundary condition, 
 and linear combination of pseudo-atomic orbital (LCPAO) coefficients, respectively. 
 It is also noted that $\langle {\bf r}\vert \phi_{\mu}^{(\bf k)}\rangle \equiv \phi_{\mu}^{(\bf k)}({\bf r})$ and
 $\langle {\bf r}\vert \chi_{{\bf R} i\alpha} \rangle \equiv \chi_{i\alpha}({\bf r}-{\bf \tau}_{i}-{\bf R})$, where 
 $i$ and $\alpha$ are atomic and orbital indices, respectively, and ${\bf \tau}_{i}$ is the position of atom $i$. 
 We assume that $M_i$ PAO functions are allocated to atom $i$. 
 Throughout the paper we do not consider the spin dependency 
 on the formulation for sake of simplicity, but the generalization is straightforward. 
 By considering overlap matrix elements 
 $ S_{{\bf R}i\alpha,{\bf R}'j\beta} \equiv  \langle \chi_{{\bf R}i\alpha}\vert \chi_{{\bf R}'j\beta} \rangle$, 
 one can introduce two alternative localized orbitals: 
 \begin{eqnarray}
    \vert \underline{\chi}_{{\bf R} i\alpha} \rangle
    &=&
    \sum_{{\bf R}' j\beta}
    \vert \chi_{{\bf R}' j\beta} \rangle
     S_{{\bf R}'j\beta, {\bf R}i\alpha}^{-1/2},\\
    \vert \widetilde{\chi}_{{\bf R} i\alpha} \rangle
    &=&
    \sum_{{\bf R}' j\beta}
    \vert \chi_{{\bf R}' j\beta} \rangle
     S_{{\bf R}'j\beta, {\bf R}i\alpha}^{-1},
 \end{eqnarray}
where $S^{-1/2}$ and $S^{-1}$ are calculated from an overlap matrix $S$ for a supercell consisting of 
$N_{\rm BC}$ primitive cells in the Born-von Karman boundary condition.
Hereafter, $\underline{\chi}$ and $\widetilde{\chi}$ will be referred to as 
L\"owdin \cite{Lowdin1950} and dual orbitals \cite{Ozaki2001}, respectively.
It is noted that we have the following relations:
 \begin{eqnarray}
  \langle \underline{\chi}_{{\bf R} i\alpha} \vert \underline{\chi}_{{\bf R}' j\beta} \rangle 
   &=& \delta_{{\bf RR}'}\delta_{ij}\delta_{\alpha\beta},\\
  \langle \chi_{{\bf R} i\alpha} \vert \widetilde{\chi}_{{\bf R}' j\beta} \rangle 
   &=& \langle \widetilde{\chi}_{{\bf R} i\alpha} \vert \chi_{{\bf R}' j\beta} \rangle = \delta_{{\bf RR}'}\delta_{ij}\delta_{\alpha\beta},
 \end{eqnarray}
and that the identity operator $\hat{I}$ can be expressed in Bloch functions $\phi$, L\"owdin orbitals $\underline{\chi}$, 
or PAOs $\chi$ and dual orbitals $\widetilde{\chi}$ as
 \begin{eqnarray}
   \nonumber
   \hat{I}
    &=&
    \sum_{{\bf k} \mu}
    \vert \phi_{\bf k \mu}\rangle
    \langle \phi_{\bf k \mu}\vert,\\
   \nonumber
    &=&
    \sum_{{\bf R} i\alpha}
    \vert \underline{\chi}_{{\bf R} i\alpha} \rangle
    \langle \underline{\chi}_{{\bf R} i\alpha} \vert,\\
    &=&
    \sum_{{\bf R} i\alpha}
    \vert \chi_{{\bf R} i\alpha} \rangle
    \langle \widetilde{\chi}_{{\bf R} i\alpha} \vert = 
    \sum_{{\bf R} i\alpha}
    \vert \widetilde{\chi}_{{\bf R} i\alpha} \rangle
    \langle \chi_{{\bf R} i\alpha} \vert.
    \label{eq:iden}
 \end{eqnarray}
 We now define a projection operator for the occupied space by 
 \begin{eqnarray}
    \hat{P}
    =
    \sum_{{\bf k} \mu}
    \vert \phi_{\bf k \mu}\rangle
     f(\varepsilon_{\bf k \mu})
    \langle \phi_{\bf k \mu}\vert,
    \label{eq:proj1}
 \end{eqnarray}
where $f$ and $\varepsilon$ are the Fermi-Dirac function and an eigenvalue of the KS equation, respectively.
By using the second line of Eq.~(\ref{eq:iden}), one obtains an alternative expression of the projection operator as 
 \begin{eqnarray}
  \hat{P}
   =
   \sum_{{\bf R} i\alpha,{\bf R}' j\beta}
    \vert \underline{\chi}_{{\bf R} i\alpha} \rangle
    \underline{\rho}_{{\bf R} i\alpha,{\bf R}' j\beta}
    \langle \underline{\chi}_{{\bf R}' j\beta} \vert
    \label{eq:proj2}
 \end{eqnarray}
 with 
 \begin{eqnarray}
   \nonumber
   \underline{\rho}_{{\bf R} i\alpha,{\bf R}' j\beta}
    &=& 
    \sum_{{\bf k} \mu}
    \langle \underline{\chi}_{{\bf R} i\alpha} \vert \phi_{\bf k \mu}\rangle
     f(\varepsilon_{\bf k \mu})
    \langle \phi_{\bf k \mu}\vert \underline{\chi}_{{\bf R}' j\beta} \rangle,\\
   \nonumber
    &=& 
    \frac{1}{N_{\rm BC}}
    \sum_{{\bf k} \mu}
    {\rm e}^{{\rm i}{\bf k}\cdot \left({\bf R}-{\bf R}'\right)}
    f(\varepsilon_{\bf k \mu})
    b_{ {\bf k} \mu,i\alpha} 
    b^{*}_{ {\bf k} \mu,j\beta},\\ 
 \end{eqnarray}
 where ${\bf b}_{{\bf k} \mu} = S^{1/2}{\bf c}_{{\bf k} \mu}$, and ${\bf c}_{{\bf k} \mu}$ is a column vector 
 whose elements are LCPAO coefficients $\{c_{ {\bf k} \mu,i\alpha}\}$. 
 As well, one can derive another expression of the projection operator by applying the third line of Eq.~(\ref{eq:iden})
 to Eq.~(\ref{eq:proj1}) as
 \begin{eqnarray}
  \hat{P}
   =
   \sum_{{\bf R} i\alpha,{\bf R}' j\beta}
    \vert \chi_{{\bf R} i\alpha} \rangle
    \rho_{{\bf R} i\alpha,{\bf R}' j\beta}
    \langle \chi_{{\bf R}' j\beta} \vert
    \label{eq:proj3}
 \end{eqnarray}
 with
 \begin{eqnarray}
   \nonumber
   \rho_{{\bf R} i\alpha,{\bf R}' j\beta}
    &=& 
    \sum_{{\bf k} \mu}
    \langle \widetilde{\chi}_{{\bf R} i\alpha} \vert \phi_{\bf k \mu}\rangle
     f(\varepsilon_{\bf k \mu})
    \langle \phi_{\bf k \mu}\vert \widetilde{\chi}_{{\bf R}' j\beta} \rangle,\\
   \nonumber
    &=&
    \frac{1}{N_{\rm BC}}
    \sum_{{\bf k} \mu}
    {\rm e}^{{\rm i}{\bf k}\cdot \left({\bf R}-{\bf R}'\right)}
    f(\varepsilon_{\bf k \mu})
    c_{ {\bf k} \mu,i\alpha} 
    c^{*}_{ {\bf k} \mu,j\beta}.\\ 
 \end{eqnarray}
 It is noted that $\underline{\rho}$ is related to $\rho$ by the following relation:
 \begin{eqnarray}
  \underline{\rho} = S^{1/2}\rho S^{1/2}.
 \end{eqnarray}
 Remembering that the number of electrons $N_{\rm ele}$ in the supercell consisting of $N_{\rm BC}$ primitive cells 
 can be obtained by the trace of $\hat{P}$, and that we have two alternative expressions Eqs.~(\ref{eq:proj2}) and (\ref{eq:proj3}) 
 for $\hat{P}$, one has the following expressions for $N_{\rm ele}$:
 \begin{eqnarray}
  \nonumber
  N_{\rm ele}
  &=&
  2{\rm tr}
  \left[
   \hat{P}
  \right],\\
  \nonumber
  &=&
  2\sum_{{\bf R} i\alpha}
  \langle \underline{\chi}_{{\bf R} i\alpha} \vert 
  \hat{P}
  \vert \underline{\chi}_{{\bf R} i\alpha} \rangle
  =
  2{\rm tr}
  \left[
   S^{1/2}\rho S^{1/2}
  \right],\\
  &=&
  2{\rm tr}
  \left[
   \rho S
  \right]
  =
  2\sum_{{\bf R} i\alpha}
  \langle \widetilde{\chi}_{{\bf R} i\alpha} \vert 
  \hat{P}
  \vert \chi_{{\bf R} i\alpha} \rangle,
 \end{eqnarray}
where the factor of 2 is due to spin degeneracy. 
Since each term in the summation over ${\bf R}$ equally contributes to $N_{\rm ele}$, 
we have $N_{\rm ele} = N_{\rm BC}N^{(\bf 0)}_{\rm ele}$, where 
$N^{(\bf 0)}_{\rm ele}= 2\sum_{i\alpha} \langle \widetilde{\chi}_{{\bf 0} i\alpha} \vert \hat{P} \vert \chi_{{\bf 0} i\alpha} \rangle$.
Thus, it is enough to consider $N^{(\bf 0)}_{\rm ele}$ instead of $N_{\rm ele}$ for further discussion. 
By introducing a notation for a subset of orbitals $\{\chi\}$ and $\{\widetilde{\chi}\}$ with 
 \begin{eqnarray}
   \vert \chi_{{\bf R}i} )
   &=& 
   \left(
     \vert \chi_{{\bf R}i1}\rangle,
     \vert \chi_{{\bf R}i2}\rangle,
     \cdots,
     \vert \chi_{{\bf R}iM_i}\rangle
   \right),\\
   \vert \widetilde{\chi}_{{\bf R}i} )
   &=& 
   \left(
     \vert \widetilde{\chi}_{{\bf R}i1}\rangle,
     \vert \widetilde{\chi}_{{\bf R}i2}\rangle,
     \cdots,
     \vert \widetilde{\chi}_{{\bf R}iM_i}\rangle
   \right),
 \end{eqnarray}
one can write $N^{(\bf 0)}_{\rm ele}$ as 
 \begin{eqnarray}
  N^{(\bf 0)}_{\rm ele} 
  = 
  2\sum_{i}
  {\rm tr}_{{\bf 0}i}
  \left[
  (\widetilde{\chi}_{{\bf 0}i}\vert 
  \hat{P} 
  \vert \chi_{{\bf 0}i})\right]
  = 
  2\sum_{i}
  {\rm tr}_{{\bf 0}i}
  \left[
   \Lambda_{{\bf 0}i}
  \right],
  \label{eq:tr_i}
 \end{eqnarray}
 where ${\rm tr}_{{\bf 0}i}$ means a partial trace over orbitals associated with an atom $i$ 
 in the central cell with ${\bf R}={\bf 0}$, and $\Lambda_{{\bf 0}i}$ is defined by 
 \begin{eqnarray}
   \Lambda_{{\bf 0}i}
   = 
   \sum_{{\bf R}j}
   \rho_{{\bf 0}i,{\bf R}j} S_{{\bf R}j,{\bf 0}i}
  \label{eq:Lambda}
 \end{eqnarray}
 with definition of block elements:
 \begin{eqnarray}
   \rho_{{\bf R}i,{\bf R}'j}
   &=& 
   (\widetilde{\chi}_{{\bf R}i}\vert \hat{P} \vert \widetilde{\chi}_{{\bf R}'j} ),\\
   S_{{\bf R}i,{\bf R}'j}
   &=& 
   (\chi_{{\bf R}i}\vert \chi_{{\bf R}'j} ).
 \end{eqnarray}
 These block elements $\rho_{{\bf R}i,{\bf R}'j}$ and $S_{{\bf R}i,{\bf R}'j}$ are $M_i\times M_j$ matrices, 
 where $M_i$ and $M_j$ are the number of PAO functions allocated to atoms $i$ and $j$, respectively.
 Therefore, $\Lambda_{{\bf 0}i}$ defined by Eq.~(\ref{eq:Lambda}) is a $M_i\times M_i$ matrix.
 It should be emphasized that Eq.~(\ref{eq:tr_i}) giving $N^{(\bf 0)}_{\rm ele}$ by the sum of the partial trace is 
 an important relation in calculating LNOs, since it shows that a {\it local} similarity transformation on 
 an atomic site $i$ does not change $N^{(\bf 0)}_{\rm ele}$ because of a property of the trace.
 Noting that $\Lambda_{{\bf 0}i}$ is nonsymmetric, we consider a general eigendecomposition of a nonsymmetric matrix
 for $\Lambda_{{\bf 0}i}$ among similarity transformations as
 \begin{eqnarray}
   V_{{\bf 0}i}^{-1}\Lambda_{{\bf 0}i}V_{{\bf 0}i} = \lambda_{{\bf 0}i},
   \label{eq:eigendecomp}
 \end{eqnarray}
 where $\lambda_{{\bf 0}i}$ is a diagonal matrix having eigenvalues $\{\lambda_{{\bf 0}i\gamma}\}$ of $\Lambda_{{\bf 0}i}$ 
 as diagonal elements.  
 Since an eigenvalue $\lambda_{{\bf 0}i\gamma}$ of $\Lambda_{{\bf 0}i}$ gives the population for 
 the corresponding eigenstate of $\Lambda_{{\bf 0}i}$, 
 one can distinguish LNOs spanning the occupied space from others among all the eigenstates of $\Lambda_{{\bf 0}i}$ 
 by monitoring the eigenvalues. To see the idea more clearly, we define an operator by 
 \begin{eqnarray}
   \hat{\Lambda}_{{\bf 0}i} 
    = 
   \sum_{\gamma}\vert v_{{\bf 0}i\gamma} \rangle\lambda_{{\bf 0}i\gamma}\langle \widetilde{v}_{{\bf 0}i\gamma}\vert,
  \label{eq:op_Lambda}
 \end{eqnarray}
 where $\vert v_{{\bf 0}i\gamma}\rangle$ is the $\gamma$-th column vector of $V_{{\bf 0}i}$, and 
 $\langle \widetilde{v}_{{\bf 0}i\gamma}\vert$ is the $\gamma$-th row vector of $V_{{\bf 0}i}^{-1}$. 
 Note that $\langle \widetilde{v}_{{\bf 0}i\gamma}\vert$ is the dual orbital of $\vert v_{{\bf 0}i\gamma}\rangle$, and 
 $\langle \widetilde{v}_{{\bf 0}i\gamma}\vert v_{{\bf 0}i\eta}\rangle = \delta_{\gamma\eta}$.
 It is easy to confirm that $\Lambda_{{\bf 0}i}$ and $\lambda_{{\bf 0}i}$ in a matrix form
 can be obtained by representing $\hat{\Lambda}_{{\bf 0}i}$ with $\{ \chi_{{\bf 0}i\alpha}\}$ and 
 $\{ \widetilde{\chi}_{{\bf 0}i\alpha}\}$, and with $\{ v_{{\bf 0}i\gamma}\}$ and $\{ \widetilde{v}_{{\bf 0}i\gamma}\}$, 
 respectively.
 Thus, we have ${\rm tr}_{{\bf 0}i}[\hat{\Lambda}_{{\bf 0}i}]={\rm tr}_{i{\bf 0}}[\Lambda_{{\bf 0}i}]$.
 If the eigenvalue $\lambda_{{\bf 0}i\gamma}$ is nearly zero, 
 the contribution of the corresponding eigenstate $\vert v_{{\bf 0}i\gamma}\rangle$ is negligible
 in the summation of Eq.~(\ref{eq:op_Lambda}). 
 Therefore, $\hat{\Lambda}_{{\bf 0}i}$ can be approximated by excluding eigenstates whose eigenvalues are less than 
 a threshold value $\lambda_{\rm th}$ that we will discuss later on. 
 The treatment can be regarded as a low-rank approximation \cite{SVD}. 
 Then, using Eqs.~(\ref{eq:tr_i}) and (\ref{eq:op_Lambda}) we can approximate the projection operator $\hat{P}$ 
 defined by Eq.~(\ref{eq:proj1}) as 
 \begin{eqnarray}
  \hat{P}
   \simeq
   \sum_{{\bf R}i}
   \sum_{\gamma}^{\lambda_{\rm th}\leq\lambda_{{\bf R}i\gamma}}
   \vert v_{{\bf R}i\gamma} \rangle\lambda_{{\bf R}i\gamma}\langle \widetilde{v}_{{\bf R}i\gamma}\vert,
   \label{eq:apop_Lambda}
 \end{eqnarray}
 where terms satisfying a condition $\lambda_{\rm th}\leq\lambda_{{\bf R}i\gamma}$ are taken into account 
 in the summation over $\gamma$. If all the terms are included, Eq.~(\ref{eq:apop_Lambda}) becomes equivalent to 
 Eq.~(\ref{eq:proj1}).  
 We now define our LNOs by $\{v\}$ whose eigenvalues are larger than or equal to the threshold value $\lambda_{\rm th}$. 
 By comparing Eq.~(\ref{eq:eigendecomp}) with Eq.~(\ref{eq:tr_i}), one has 
 $V_{{\bf 0}i}^{-1}\Lambda_{{\bf 0}i}V_{{\bf 0}i}=
 V_{{\bf 0}i}^{-1}(\widetilde{\chi}_{{\bf 0}i}\vert \hat{P}\vert \chi_{{\bf 0}i})V_{{\bf 0}i}$. 
 Thus, LNOs are defined by $\{v\}$, and it should be noted that $\{\widetilde{v}\}$ are the corresponding dual orbitals.  
 The approximate formula Eq.~(\ref{eq:apop_Lambda}) for $\hat{P}$ implies that a set of orbitals $\{v\}$, LNOs, 
 well span the occupied space, while the number of orbitals is reduced compared to the original PAOs.
 It is worth noting that LNOs can be independently calculated for each atom by a non-iterative calculation
 via Eqs.~(\ref{eq:Lambda}) and (\ref{eq:eigendecomp}). 
 The fact makes the method of generating LNOs very efficient, and also guarantees 
 that the resultant LNOs associated with an atom $i$ are expressed by a linear combination of PAOs allocated 
 to only the atom $i$ \cite{multi-site-orbitals}. 

 The computational procedure to generate LNOs is summarized as follows:
 (i) Calculation of $\Lambda$. 
 For each atom $i$ the matrix $\Lambda_{{\bf 0}i}$ is calculated by Eq.~(\ref{eq:Lambda}), 
 where the summation over ${\bf R}$ is limited within a finite range because of the locality of PAOs in real space.
 (ii) Diagonalization of $\Lambda$. 
 Since the matrix $\Lambda_{{\bf 0}i}$ is nonsymmetric, the diagonalization in 
 Eq.~(\ref{eq:eigendecomp}) is performed  by a generalized eigenvalue solver for a nonsymmetric matrix 
 such as DGEEV in LAPACK \cite{LAPACK}.
 (iii) Selection of $v$. 
 Eigenvectors $\{v_{{\bf 0}i\gamma}\}$ whose eigenvalues are larger than or equal to the threshold 
 value $\lambda_{\rm th}$ are selected as LNOs.
 \noindent
 Only the overlap and density matrices are required to calculate LNOs through the steps (i)-(iii) above. 
 Therefore, either conventional O($N^3$) methods 
 or O($N$) methods can be employed as eigenvalue solver as long as they generate the density matrix. 
 As for LNOs other than those in the central cell with ${\bf R}={\bf 0}$, it is apparent from the derivation that 
 one can obtain $\vert v_{{\bf R}i\gamma}\rangle$ by parallel translation of $\vert v_{{\bf 0}i\gamma}\rangle$ 
 with the lattice vector ${\bf R}$. 
 The method can also be extended to choose another energy window. 
 In the projection operator $\hat{P}$ defined by Eq.~(\ref{eq:proj1}), the Fermi-Dirac function is introduced
 to choose the occupied space. However, one can choose a proper energy window in the definition of 
 the projection operator $\hat{P}$ depending on what we discuss, which enables us to focus on specific bands 
 such as localized $d$-bands near the Fermi level. In this sense, LNOs can be utilized like 
 Wannier functions (WFs) \cite{Marzari1997,Marzari2012}, 
 while WFs are obtained through a unitary transformation of Bloch functions rather than the low-rank approximation, 
 and they are orthonormal each other unlike LNOs. It is also worth pointing out that our method is similar to 
 the quasiatomic orbitals schemes based on maximizing a similarity measure calculated by a projection \cite{Lu2004,Chan2007,Qian2008}.

 It might be possible for the method we present in the paper to be applied for other localized basis functions 
 such as finite element methods \cite{Tsuchida1996,Fattebert2007} and finite difference methods \cite{Chelikowsky1994,Iwata2010,Ono2005}. 
 In those cases one may introduce spatial partitioning methods such as Voronoi tessellation to decompose basis functions 
 rather than focusing on basis functions on a single grid. A similar procedure can be applied for a set of partitioned basis functions, 

\subsection{DC method with LNOs}

 Here we consider an extension of the DC method \cite{Yang1991,Yang1995} using LNOs discussed in the previous subsection.
 Our theoretical basis to formulate the O($N$) DC method is that the total energy and atomic forces 
 in the KS framework can be calculated by using electron density $n({\bf r})$, density matrix $\rho$, 
 and energy density matrix $e$ defined by 
 \begin{eqnarray}
    \nonumber
    n({\bf r})
     &=& 
    \sum_{i}
    \left(
      2\sum_{\alpha,{\bf R}j\beta}
      \rho_{{\bf 0}i\alpha,{\bf R}j\beta}\chi_{{\bf 0}i\alpha}({\bf r})\chi_{{\bf R}j\beta}({\bf r})    
    \right),\\
     &=& 
    \sum_{i}n_i({\bf r}), 
   \label{eq:ed}
 \end{eqnarray}
 \begin{eqnarray}
    \rho 
     = 
    -\frac{1}{\pi}{\rm Im}
   \int_{-\infty}^{\infty} G(E+i0^+)f(E)dE,  
   \label{eq:dm}
 \end{eqnarray}
 and 
 \begin{eqnarray}
    e 
     = 
    -\frac{1}{\pi}{\rm Im}
   \int_{-\infty}^{\infty} G(E+i0^+)f(E)E dE,  
   \label{eq:edm}
 \end{eqnarray}
 where $G$ is the Green function defined by $G(Z)\equiv (ZS-H)^{-1}$ with the overlap matrix $S$ 
 and KS matrix $H$, and the factor of 2 in Eq.~(\ref{eq:ed}) is due to spin degeneracy.
 It is remarked that forces on atoms in the DC method are not calculated variationally, but evaluated 
 by using the formula derived by assuming that numerically exact KS wave functions are available as discussed 
 in Ref.~\cite{Ozaki2006}. 
 The DC method calculates the Green function $G(Z)$ approximately by introducing 
 the truncation scheme as shown in Fig.~1. The KS matrix of the truncated cluster 
 for atom $i$ are constructed using PAOs and LNOs as follows:
 \begin{eqnarray}
    H^{(i)} =
     \left(
       \begin{array}{c|c}  
         {\rm PAO-PAO} & {\rm PAO-LNO}\\
          \hline
         {\rm LNO-PAO} & {\rm LNO-LNO}\\
       \end{array}  
     \right),
   \label{eq:Hi}
 \end{eqnarray}
 where the top left and bottom right blocks correspond to the short range (orange) region
 represented by PAOs, and the long range (yellow) region represented by LNOs, respectively, as shown in Fig.~1.
 The top right and the bottom left block consist of the hopping matrix elements 
 bridging the two regions, and they are represented by both PAOs and LNOs. 
 As well, the same structure is found for the overlap matrix.  
 Noting that the computational bottleneck is mainly governed by the eigenvalue problem 
 for the truncated clusters, and that the matrix size can be reduced by introducing LNOs
 compared to the conventional DC method, one can expect a considerable reduction of the 
 computational cost as the size of the long range region increases. 
 The idea of reducing the matrix dimension by introducing an effective representation of Hamiltonian 
 is similar to that in the O($N$) Krylov subspace method \cite{Ozaki2006} and the absolutely localized molecular 
 orbitals (ALMO) method \cite{Khaliullin2013}.
 By solving the eigenvalue problem $H^{(i)}c^{(i)}_{\mu}=\varepsilon^{(i)}_{\mu}S^{(i)}c^{(i)}_{\mu}$
 for the truncated cluster of atom $i$, we calculate matrix elements associated with the atom $i$ for 
 the Green function as \cite{DC-multicore}
 \begin{eqnarray}
    G^{(i)}_{{\bf 0}i\alpha,{\bf R}j\beta}(Z)
          = \sum_{\mu}
             \frac{c^{(i)}_{\mu,{\bf 0}i\alpha}(c^{(i)}_{\mu,{\bf R}j\beta})^*}
                  {Z-\varepsilon^{(i)}_{\mu}}.
   \label{eq:Gi}
 \end{eqnarray}
 Matrix elements $\rho_{{\bf 0}i\alpha, {\bf R}j\beta}$ and $e_{{\bf 0}i\alpha, {\bf R}j\beta}$ associated 
 with the atom $i$ can be calculated by inserting Eq.~(\ref{eq:Gi}) into Eqs.~(\ref{eq:dm}) and (\ref{eq:edm}),
 respectively. 
 We only have to calculate $\rho_{{\bf 0}i\alpha, {\bf R}j\beta}$ and $e_{{\bf 0}i\alpha, {\bf R}j\beta}$
 only if $S_{{\bf 0}i\alpha,{\bf R}j\beta}$ is non zero, since the other elements do not contribute to  
 the total energy and forces on atoms in case of semi-local functionals such as 
 local density approximations (LDA) \cite{Ceperley1980,Perdew1981}
 and generalized gradient approximations (GGA) \cite{Perdew1996}. 
 Then, $n_i({\bf r})$ in Eq.~(\ref{eq:ed}) is easily computed from $\rho_{{\bf 0}i\alpha, {\bf R}j\beta}$.
 By applying the procedure for all the atoms in a system, all the necessary information to calculate 
 the total energy and forces on atoms are obtained. 

 The overall procedure of the DC method with LNOs is summarized as follows:
 (i) Calculation of LNOs. LNOs are calculated by Eq.~(\ref{eq:eigendecomp}) for all atoms in the central 
 cell with ${\bf R}={\bf 0}$. At every SCF step LNOs are updated, leading to self-consistent determination of LNOs. 
 (ii) Construction of $H^{(i)}$ and $S^{(i)}$.
 For each atom $i$ the KS and overlap matrices for a truncated cluster associated with the atom $i$
 are constructed by Eq.~(\ref{eq:Hi}).
 (iii) Diagonalization of $H^{(i)}c^{(i)}_{\mu}=\varepsilon^{(i)}_{\mu}S^{(i)}c^{(i)}_{\mu}$ 
 by making use of a parallel eigenvalue solver.  
 (iv) Finding a common chemical potential to conserve the total number of electrons in the system using Eq.~(\ref{eq:Gi}), 
 as discussed in Ref.~\cite{Ozaki2006} in detail.
 (v) Calculation of density matrix $\rho$ and energy density matrix $e$ using Eqs.~(\ref{eq:dm}), (\ref{eq:edm}), 
 and (\ref{eq:Gi}).
 (vi) Calculation of electron density $n$ using Eq.~(\ref{eq:ed}).

\section{IMPLEMENTATIONS}

We have implemented the DC-LNO method into the OpenMX DFT software package \cite{OpenMX} which is based on norm-conserving 
pseudopotentials (PPs) \cite{MBK1993,Theurich2001} and optimized pseudo-atomic orbitals (PAOs) \cite{Ozaki2003,Ozaki2004} as basis set. 
All the benchmark calculations were performed with a computational condition of a production level. 
The basis functions used are C6.0-s2p2d1, Si7.0-s2p2d1, Ti7.0-s2p2d1, O6.0-s2p2d1, Li8.0-s3p2, Al7.0-s2p2d1, and 
Fe5.5-s3p2d2 for carbon, silicon, titanium, oxygen, lithium, aluminum, and iron, respectively, 
where in the abbreviation of basis functions such as C6.0-s2p2d1, C stands for the atomic symbol, 
6.0 the cutoff radius (Bohr) in the generation by the confinement scheme, and s2p2d1 means the employment 
of two, two, and one optimized radial functions for the $s$-, $p$-, and $d$-orbitals, respectively.
The radial functions were optimized by a variational optimization method \cite{Ozaki2003}. 
As valence electrons in the PPs we included $2s$ and $2p$, $3s$ and $3p$, $3s$, $3p$, $3d$, and $4s$, 
$2s$ and $2p$, $1s$, $2s$, and $2p$, 
$3s$ and $3p$, and $3s$, $3p$, $3d$, and $4s$ states for carbon, silicon, titanium, oxygen, lithium, aluminum, 
and iron, respectively. 
All the PPs and PAOs we used in the study were taken from the database (2013) in the OpenMX website \cite{OpenMX}, which 
were benchmarked by the delta gauge method \cite{Lejaeghere2016}. Real space grid techniques are used for the numerical integrations 
and the solution of the Poisson equation using FFT with the energy cutoff of 300 Ryd \cite{Ozaki2005}.
We used a generalized gradient approximation (GGA) proposed by Perdew, Burke, and Ernzerhof
to the exchange-correlation functional \cite{Perdew1996}. An electronic temperature of 300 K is used to count the number
of electrons by the Fermi-Dirac function for all the systems we considered.

\begin{figure}[t]
    \centering
    \includegraphics[width=8.7cm]{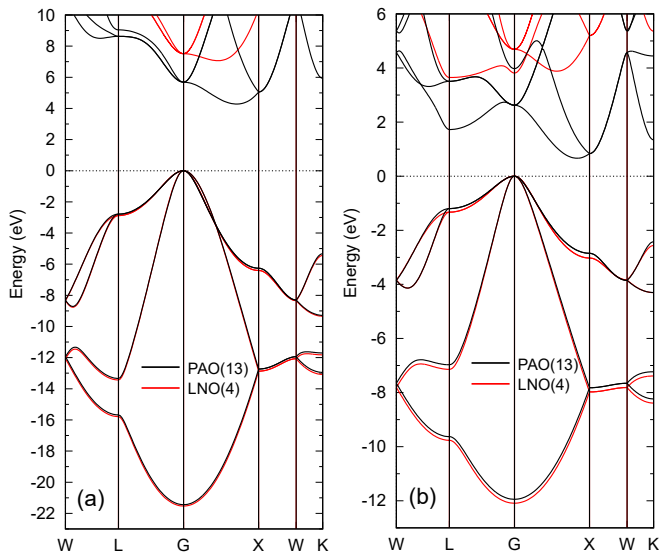}
    \caption{Band dispersions of (a) diamond and (b) silicon in the diamond structure with experimental lattice constants 
             (3.567 and 5.430~\AA) calculated by PAOs (black) and LNOs (red). 
             A conventional O($N^3$) method was used for the diagonalization, where 
             the number of {\bf k}-points for the Brillouin zone sampling is $71\times 71\times 71$ for both the cases.
             In the case of LNOs, the SCF calculations were performed by using PAOs, and after determining the SC 
             electron density, LNOs were used to calculate the band dispersion. The numbers of PAOs and LNOs 
             per atom are shown in the parenthesis.
             }
\end{figure}

\begin{figure}[t]
    \centering
    \includegraphics[width=8.7cm]{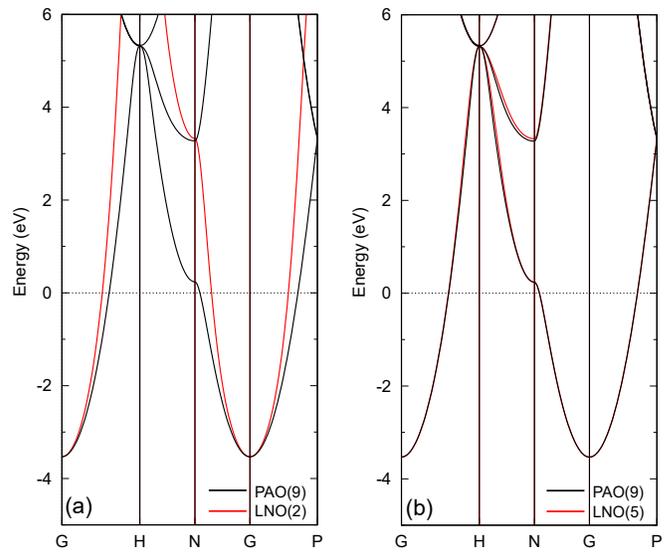}
    \caption{Band dispersions of BCC lithium with an experimental lattice constant 
             of 3.491~\AA~calculated by PAOs (black) and (a) two and (b) five LNOs (red).
             The number of {\bf k}-points for the Brillouin zone sampling is $101\times 101\times 101$.
             The other details are the same as in the caption of Fig.~2.
             }
\end{figure}

The short and long range regions depicted in Fig.~1 are determined as follows: 
(i) We first pick up atoms in a sphere with a given cutoff radius $r_{\rm L}$. 
(ii) Among the atoms selected by the step (i) we distinguish the first neighboring atoms (FNAs) 
having non-zero overlap with the central atom in terms of basis functions, and 
remaining atoms other than FNAs are called the second neighboring atoms (SNAs), where 
the number of FNAs and SNAs are $N_{\rm F}$ and $N_{\rm S}$, respectively. 
(iii) The short range region is determined by adjusting a cutoff radius $r_{\rm S}$
so that the number of atoms in a sphere with a radius of $r_{\rm S}$ can be 
as close as possible to $N_{\rm F}+\kappa N_{\rm S}$, 
where the parameter $\kappa$ can vary from 0 to 1, and 
we will discuss the choice of $\kappa$ later on.
(iv) The long range region consists of 
remaining atoms other than atoms selected by the step (iii). 
If we assign FNAs to atoms in the short range region, the total energy does not converge to 
the numerically exact one calculated by the conventional diagonalization method even if 
the cutoff radius $r_{\rm L}$ increases systematically. This is because the error 
with the low-rank approximation by Eq.~(\ref{eq:apop_Lambda}) keeps increasing 
as the cutoff radius increases. To avoid the situation, we add a buffer region consisting  
of about $\kappa N_{\rm S}$ atoms as described by the step (iii) above, which guarantees
the convergence of the total energy and other quantities as a function of the cutoff radius $r_{\rm L}$.
Throughout the study, we used $\kappa$ of $\frac{3}{10}$ for all the systems by taking the accuracy 
into account more than the efficiency, and did not adjust the parameter, while a smaller value, 
which well balances both the accuracy and efficiency, can be employed for some systems. 

The way of parallelization for the DC-LNO method on parallel computers will be discussed
together with its benchmark calculations later on.

\section{NUMERICAL RESULTS}

\subsection{Band dispersions by LNOs}

In order to investigate to what extent LNOs can span occupied spaces, we compare band dispersions 
of gapped and metallic systems calculated with PAOs and LNOs. Figure~2(a) and (b) show 
band dispersions of diamond and silicon calculated by a conventional O($N^3$) diagonalization method 
with PAOs and LNOs. For both the cases the SCF calculations were performed by using PAOs. 
   \begin{table}[b]
    \caption{
     Eigenvalues $\lambda$ of the matrix $\Lambda$ for diamond, silicon, BCC lithium, and FCC aluminum.
     The corresponding eigenvectors were used as LNOs to calculate the band dispersions 
     shown in Figs.~2-4.
    }
   \vspace{1mm}
   \begin{tabular}{lcccccccccccc}
   \hline\hline
                     &&& Diamond  &&&  Si  &&&  Li  &&& Al\\
     \hline
      $\lambda_1$    &&&   0.514  &&& 0.639&&& 0.999&&& 0.551 \\
      $\lambda_2$    &&&   0.483  &&& 0.430&&& 0.277&&& 0.253 \\
      $\lambda_3$    &&&   0.483  &&& 0.430&&& 0.075&&& 0.253 \\
      $\lambda_4$    &&&   0.483  &&& 0.430&&& 0.075&&& 0.253 \\
      $\lambda_5$    &&&   0.012  &&& 0.018&&& 0.074&&& 0.049 \\
      $\lambda_6$    &&&   0.012  &&& 0.018&&& 0.001&&& 0.049 \\
      $\lambda_7$    &&&   0.012  &&& 0.018&&& 0.001&&& 0.049 \\
      $\lambda_8$    &&&   0.011  &&& 0.012&&& 0.001&&& 0.033 \\
      $\lambda_9$    &&&   0.011  &&& 0.012&&&-0.002&&& 0.033 \\
      $\lambda_{10}$ &&&  -0.001  &&& 0.002&&&   -  &&&-0.002 \\
      $\lambda_{11}$ &&&  -0.001  &&& 0.002&&&   -  &&&-0.002 \\
      $\lambda_{12}$ &&&  -0.001  &&& 0.002&&&   -  &&&-0.002 \\
      $\lambda_{13}$ &&&  -0.018  &&&-0.011&&&   -  &&&-0.015 \\
   \hline
   \end{tabular}
  \end{table}
For the case of LNOs the band dispersions were calculated with the LNOs after the SCF calculations 
with PAOs. The number of LNOs per atom is 4 for both carbon and silicon atoms. 
It is found that in both the cases the band dispersions of occupied space are well reproduced 
with LNOs compared to those calculated by PAOs, while a large difference can be seen in conduction bands
between PAOs and LNOs as expected. The good agreement between PAOs and LNOs in describing 
the occupied bands implies that the low-rank approximation by Eq.~(\ref{eq:apop_Lambda}) 
is reasonably valid. As shown in Table I we see that the first four eigenvalues of the matrix $\Lambda$ 
are actually dominant for both diamond and silicon, justifying the low-rank approximation.
The largest and the next three eigenvalues correspond to a $s$-orbital and $p$-like orbitals 
deformed by contribution of $d$-orbitals, respectively. 
It is also noted that the eigenvalues can be negative, which is related to a negative value 
of Mulliken populations for delocalized orbitals \cite{Whangbo1978, Gomez2009}. 
As well as the gapped systems, similar calculations were performed for metals, lithium 
in the body centered cubic (BCC) structure and aluminum in the face centered cubic (FCC) structure 
as shown in Figs.~3 and 4, respectively. 
The band dispersions calculated with the minimal LNOs are reasonably compared to those by PAOs, 
while the use of the five and nine LNOs for Li and Al atoms fully reproduce the band dispersions including conduction 
bands as shown in Figs.~3(b) and 4(b), respectively. 
One can confirm again in Table I
that eigenvalues for the minimal LNOs are dominant even for metals, while the magnitude of 
the subsequent eigenvalues is relatively large compared to those of the gapped systems. 
Thus, we conclude that LNOs can be regarded as a compact basis set spanning well the occupied space 
for both gapped and metallic systems. 

\begin{figure}[t]
    \centering
    \includegraphics[width=8.7cm]{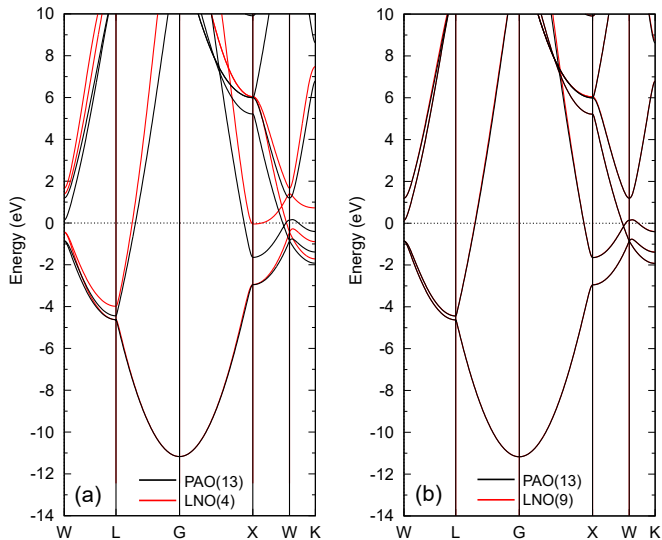}
    \caption{Band dispersions of FCC aluminum with an experimental lattice constant 
             of 4.050~\AA~calculated by PAOs (black) and (a) four and (b) nine LNOs (red).
             The number of {\bf k}-points for the Brillouin zone sampling is $111\times 111\times 111$.
             The other details are the same as in the caption of Fig.~2. 
             }
\end{figure}

\begin{figure*}[ht]
\begin{center}
    \includegraphics[width=15.5cm]{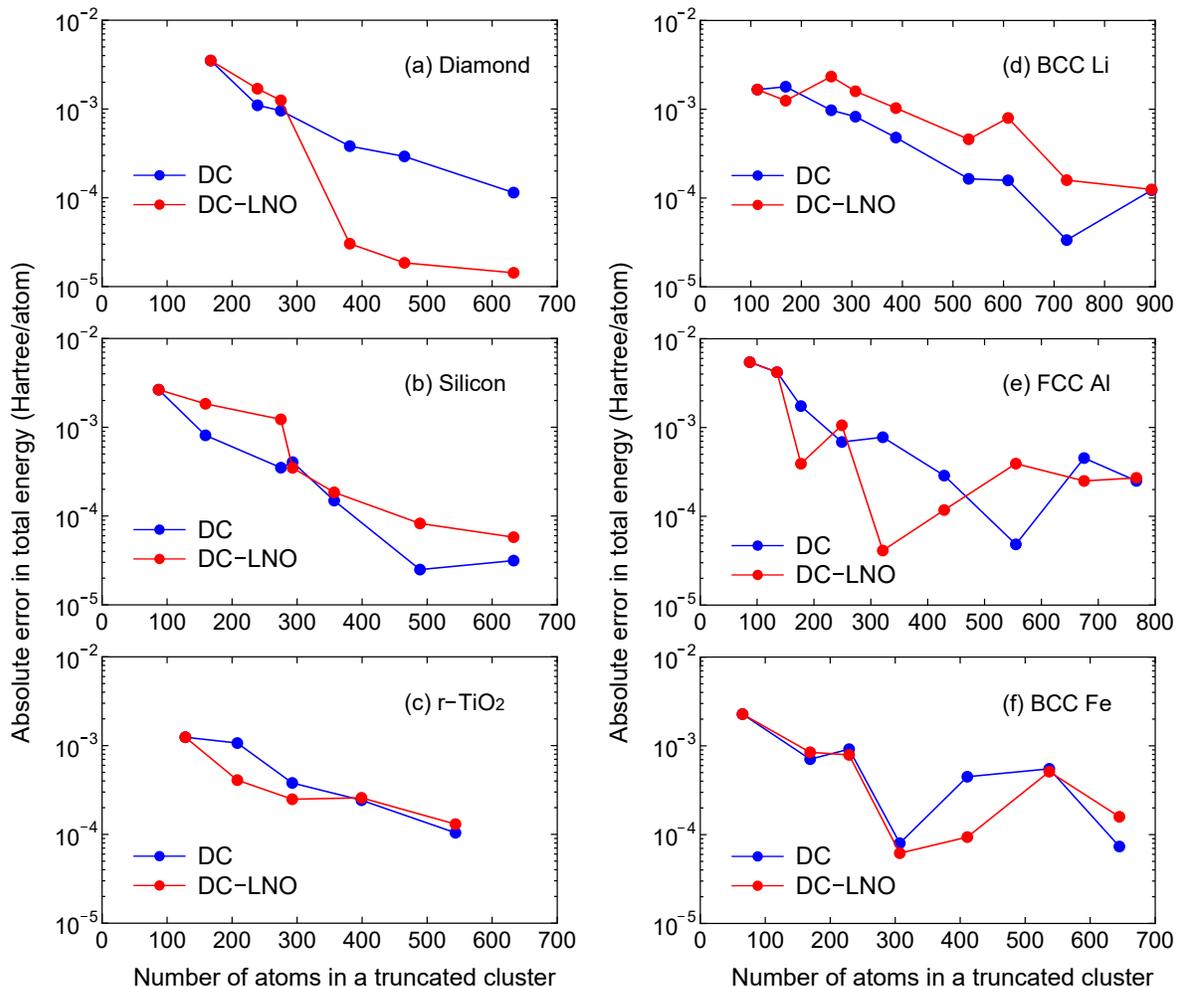}
    \caption{Absolute error in the total energy (Hatree/atom) for (a) diamond, (b) silicon in the diamond structure, 
    (c) rutile TiO$_2$, (d) BCC lithium, (e) FCC aluminum, and (f) BCC iron as a function of the number of atoms 
    in a truncated cluster calculated by the DC and DC-LNO methods. The experimental lattice constants were used 
    for all the cases.
             }
\end{center}
\end{figure*}

\subsection{Total energies by DC-LNO}

As a first step of validation of the DC-LNO method with respect to computational accuracy and efficiency, 
we show in Fig.~5 the absolute error in the total energy of gapped and metallic systems calculated 
by the DC and the DC-LNO methods, where the reference energies were calculated by the conventional 
O($N^3$) method with dense {\bf k}-points for the Brillouin zone sampling as given in the caption of Figs.~2-4.
For all the cases the threshold value $\lambda_{\rm th}$ in Eq.~(\ref{eq:apop_Lambda}) was set to be 0.1 
which gives us the minimal LNOs corresponding to orbitals of valence electrons. 
In the gapped systems, diamond, silicon, and rutile TiO$_2$, the absolute error decreases almost 
exponentially as the number of atoms in a truncated cluster increases. The overall behaviors of 
the error between the DC and DC-LNO methods are similar to each other, while the error 
by the DC-LNO method is accidentally much smaller than that by the DC method in the case of large truncated 
clusters of diamond. As well as the gapped systems the absolute error for metals calculated by the DC-LNO 
method decreases as increasing the number of atoms in a truncated cluster in a similar way to the DC method. 
The relatively large oscillating behavior observed in the metals might be related to long range characteristics 
of the off-diagonal Green functions as discussed in the Appendix. 
For all the cases including metals it is found that a truncated cluster including about 300 atoms is required 
to attain the millihartree accuracy corresponding to the error less than a few millihartree/atom in the total energy.
From the comparison between the DC and DC-LNO methods, we see that the computational accuracy does not degrade largely 
even if basis functions for atoms in the long range region are approximated by LNOs, and that thereby the computational 
accuracy can be controlled mainly by the size of truncated cluster just like for the DC method. 
Since controlling only the single parameter allows us to balance the computational accuracy and efficiency, 
it is expected that the feature makes the DC-LNO method easy to use for a wide variety of applications. 

\begin{figure}[t]
    \includegraphics[width=8.7cm]{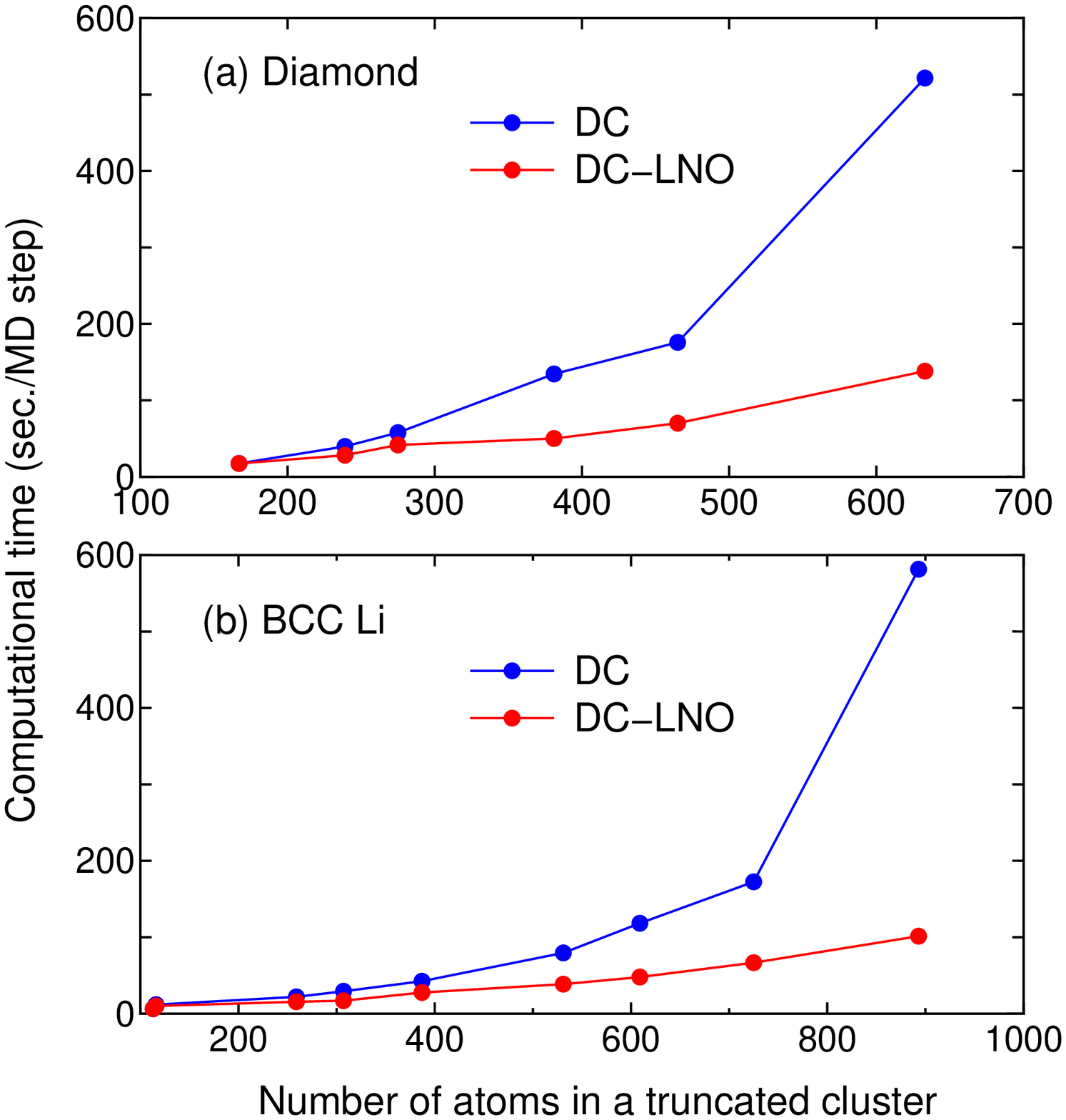}
    \caption{Comparison of computational time of the diagonalization part per molecular dynamics (MD) step 
             between the DC and DC-LNO methods for (a) diamond and (b) BCC lithium. 
             Both the calculations were performed for the primitive cell using 14 MPI processes per atom 
             on Intel Xeon CPUs E5-2690v4 @ 2.60~GHz. 
             }
\end{figure}

\subsection{Computational time}

Since the total number of basis functions to represent the Hamiltonian of the truncated cluster is 
reduced by introducing LNOs while keeping the accuracy, 
it is expected that the computational time can be substantially reduced. 
In the whole procedure of the DC-LNO method the calculation of LNOs and construction of Hamiltonian 
and overlap matrices occupy a small fraction of the whole computational time, typically less than $10\%$, 
and thereby the computational time is mainly governed by solving of the eigenvalue problem 
$H^{(i)}c^{(i)}_{\mu}=\varepsilon^{(i)}_{\mu}S^{(i)}c^{(i)}_{\mu}$ for each atom $i$. 
Noting that the computational time to solve the eigenvalue problem scales as the third power 
of the dimension of the matrices, the ratio of computational time between the DC-LNO and DC methods 
for elemental systems might be estimated by 
 \begin{eqnarray}
    \frac{t_{\rm DC-LNO}}{t_{\rm DC}}
    = 
    \frac{\left[ M_{\rm P}(N_{\rm F}+\kappa N_{\rm S})+M_{\rm L}(1-\kappa)N_{\rm S}\right]^3}
         {M_{\rm P}^3(N_{\rm F}+N_{\rm S})^3},
   \label{eq:time-ratio}
 \end{eqnarray}
where $M_{\rm P}$ and $M_{\rm L}$ are the number of PAOs and LNOs associated with each atom, respectively,
and $\kappa$ is a factor, which is fixed to $\frac{3}{10}$ in this study, to control the size of the buffer region
as discussed in the section 'IMPLEMENTATIONS'.
For example, if the cutoff radius $r_{\rm L}$ is set to be $8.7$~\AA~in the diamond structure with 
the experimental lattice constant of 3.567~\AA, $N_{\rm F}$ and $N_{\rm S}$ are found to be 167 and 298, 
and the resultant number of atoms in the short (long) range region becomes 275 (190).
Then, $t_{\rm DC-LNO}/t_{\rm DC}$ can be estimated to be 0.37 in the case that the number of PAOs
and LNOs per atom is 13 and 4, which implies that the computational time of the DC-LNO method becomes
about one third of that calculated by the DC method. 
Figure 6 shows actual timing results of the DC and DC-LNO methods for (a) diamond and (b) BCC lithium. 
We see that the actual $t_{\rm DC-LNO}/t_{\rm DC}$ for the diamond case is 0.40 in the case that the number 
of atoms is 465 (=167+298) as shown in Fig.~6(a), which is well compared to the estimated value of 0.37. 
As indicated by Eq.~(\ref{eq:time-ratio}) and in Figs.~6(a) and (b), it is concluded that the DC-LNO method 
becomes much faster than the DC method as the size of the truncated cluster increases. 

\subsection{Parallelization}

To minimize the computational time on massively parallel computers we introduce a multilevel parallelization 
using message passing interface (MPI). In our implementation there are three levels for the parallelization, 
i.e., atom level, spin level, and diagonalization level as explained below:
(i) {\it Parallelization in the atom level}.
If the number of MPI processes is smaller than that of atoms,
only the parallelization in the atom level is taken into account. 
The allocation of atoms to MPI processes is performed by a bisection method which is 
based on a projection of atoms onto a principal axis calculated from an inertia tensor and a modified binary 
tree of MPI processes to minimize memory usage and amount of MPI communications \cite{Duy2014}. 
(ii) {\it Parallelization in the spin level}.
If the number of MPI processes exceeds that of atoms, and the spin polarized calculation is performed, 
the parallelization in the spin level is introduced on top of the parallelization in the atom level, where 
a loop for the spin index is further parallelized. 
(iii) {\it Parallelization in the diagonalization level}.
If the number of MPI processes is larger than the product of the number of atoms and the multiplicity of spin index,
corresponding to 1, 2, and 1 for non-spin polarized, spin-polarized, and non-collinear calculations, respectively,
a parallelization in the diagonalization level is further taken into account on top of both the parallelizations 
in the atom and spin levels. The parallelization in the diagonalization level is made by employing a parallel eigenvalue 
solver ELPA \cite{ELPA}. It is noted that the parallelization in the diagonalization level requires a considerable amount 
of MPI communications, while the parallelizations in the atom and spin levels have less MPI communications. 
So, one would expect a high parallel efficiency in the atom and spin levels, while the parallelization 
in the diagonalization level might be limited up to several tens of MPI processes. 
To achieve a better scaling for the parallelization in the diagonalization level, it is important to allocate 
CPU cores in the same computer node as MPI processes to avoid the inter-node communication as much as possible. 
We have implemented the multilevel parallelization so that amount of the inter-node communication can be minimized 
especially for the parallelization in the diagonalization level. 
In Fig.~7 the speed-up ratio in the MPI parallelization of the DC-LNO method is shown for 
non-spin polarized calculations of a diamond supercell containing 64 atoms. 
Since the multiplicity of spin index is 1, we see a nearly ideal behavior up to 64 MPI processes. 
Beyond 64 MPI processes the parallelization in the diagonalization level is taken into account on top of 
the parallelization in the atom level. 
A superlinear speed-up is observed at 128 and 256 MPI processes, which might be due to an effective use 
of cache by the reduction of memory usage, and a good scaling is achieved up to 1280 MPI processes at which 
the parallel efficiency is calculated to be 70\% using the elapsed time at 1 MPI process as reference.
Since each computer node has 20 CPU cores in this case, it would be reasonable to observe the good scaling up to 
1280 (=64$\times$20) MPI processes. 
Thus, we see that the multilevel parallelization is very effective to 
minimize the computational time in accordance with a recent development of massively parallel computers.

\begin{figure}[t]
    \includegraphics[width=8.7cm]{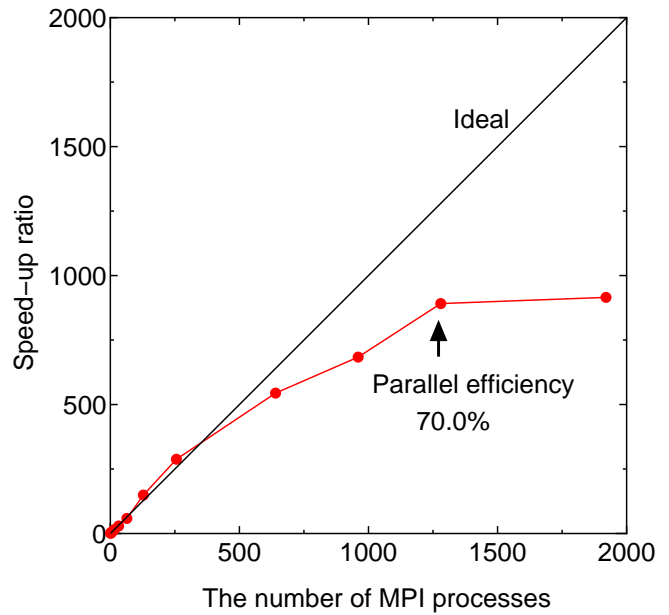}
    \caption{Speed-up ratio in the MPI parallelization of the DC-LNO method for a diamond supercell 
             containing 64 atoms, where the cutoff radius $r_{\rm L}$ of $8.0$~\AA~was used, leading to 
             the numbers of atoms of 239 and 142 in the short and long range regions, and the dimension of matrices
             of 3675 for the truncated cluster problem.
             The calculations were performed using a cluster machine consisting 
             of Intel Xeon CPUs E5-2680v2 @ 2.80~GHz connected with Infiniband 4X FDR @ 56~Gbps.
             }
\end{figure}

\begin{figure*}[ht]
\begin{center}
    \includegraphics[width=16.7cm]{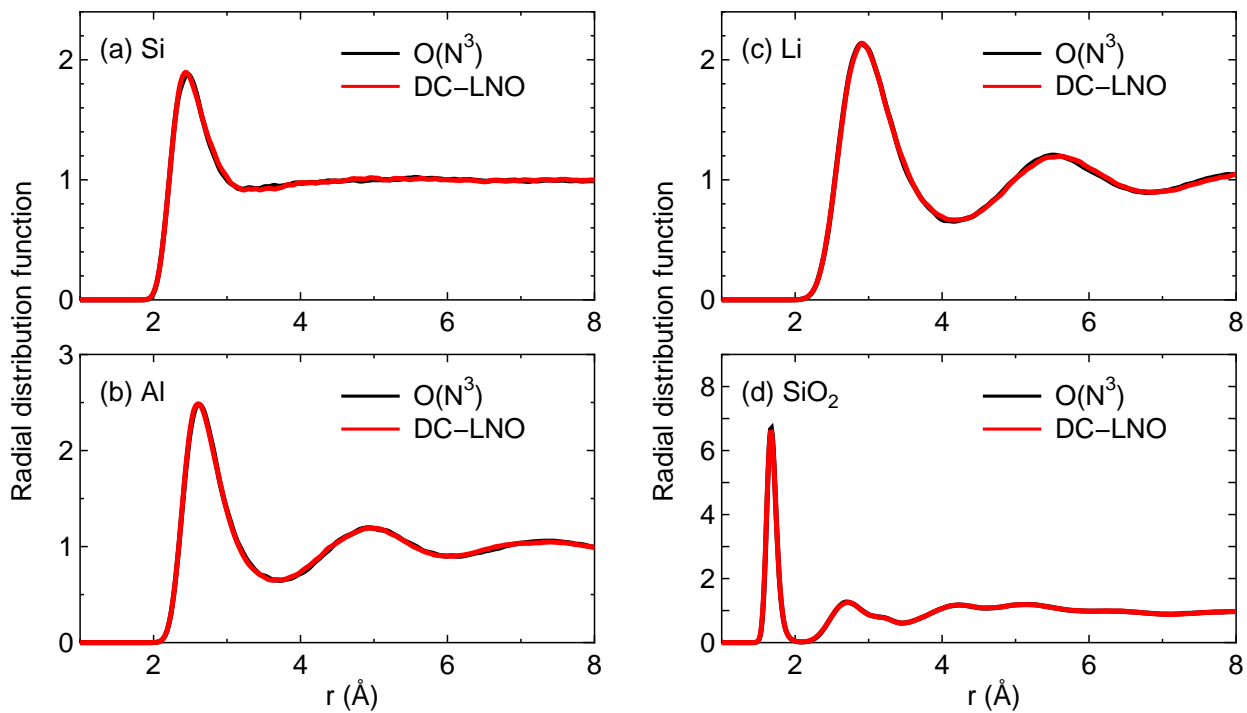}
    \caption{Total radial distribution function (RDF) of (a) silicon at 3500~K, (b) aluminum at 2500~K, 
             (c) lithium at 800~K, and (d) SiO$_2$ at 3000~K, calculated by the conventional $O(N^3)$ diagonalization 
             and the DC-LNO methods. 
             The MD simulations were performed for cubic supercells containing 64, 108, 128 and 192 atoms 
             with a fixed lattice constant of 10.86, 12.15, 14.04, 14.25 \AA~for silicon, aluminum, lithium, 
             and SiO$_2$, respectively, for 10 ps with the time step of 2 fs. The temperature was controlled 
             by a velocity scaling scheme by Woodkock \cite{Woodcock}.
             The coordinates for the first 1 ps were excluded to calculate RDF. 
             In the DC-LNO method the cutoff radii $r_{\rm L}$ of 11.3, 10.1, 12.5, and 11.0 were used for silicon, 
             aluminum, lithium, and SiO$_2$, respectively, resulting in truncated clusters consisting of 293, 249, 339, 
             and 344 (on average) atoms in the ideal bulk structures. In the conventional $O(N^3)$ diagonalization method, 
             {\bf k}-points of $7\times 7\times 7$, $8\times 8\times 8$, $7\times 7\times 7$, and $5\times 5\times 5$
             were used for the Brillouin zone sampling in silicon, aluminum, lithium, and SiO$_2$, respectively,
              }
\end{center}
\end{figure*}

\subsection{Molecular dynamics simulations}

To demonstrate the applicability of the DC-LNO method for molecular dynamics (MD) simulations, 
we show radial distribution functions (RDFs) in liquid phases of silicon, aluminum, lithium, and SiO$_2$ in Fig.~8.
Since the electronic structures exhibit metallic features in the liquid phases of silicon, aluminum, and lithium, 
the MD simulations can be considered as a severe benchmark to validate the applicability of the DC-LNO method to metals. 
The cutoff radii $r_{\rm L}$ we used correspond to truncated clusters consisting of about 300 atoms in the ideal bulk structures.
It turns out that in all the cases the DC-LNO method reproduces well the results by the conventional $O(N^3)$ diagonalization method, 
and that the obtained RDFs are well compared to other computational results \cite{Behler2007,Recoules2005,Anta1999,Kim2012}.
The considerable agreement between the DC-LNO and conventional methods strongly implies that a sufficient accuracy in reproducing 
at least RDF for MD simulations can be attainable with a cutoff radius $r_{\rm L}$ resulting in truncated clusters consisting 
of about 300 atoms for not only insulators but also metals. Thus, adjusting the cutoff radius $r_{\rm L}$ so that the number 
of atoms in a truncated cluster can be $\sim$ 300 atoms would be a compromise to balance the computational accuracy 
and efficiency, while the difference between the DC and DC-LNO methods in terms of the computational efficiency 
may not be significant for truncated clusters of this size. 
It is crucial to minimize the elapsed time for realization of long time MD simulations. 
With the computational condition the elapsed time per SCF step for silicon is 1.5 (sec.) on average
using 1280 MPI processes on the same machine used for the calculations shown in Fig.~7.

\section{CONCLUSIONS}

We have presented an efficient O($N$) method based on the DC approach and a coarse graining of basis functions by 
localized natural orbitals (LNOs) for large-scale DFT calculations. A straightforward way to attain sufficient accuracy 
in the DC method is to employ a relatively large cutoff radius for the truncation of a system, which is the most fundamental 
parameter in most of O($N$) methods to control the computational accuracy and efficiency. 
We have adopted the rather brute force approach, and attempted to decrease the computational cost by introducing LNOs 
as basis functions in the long range region of the truncated cluster, and to minimize the elapsed time in the computation 
with the help of a multilevel parallelization. 
The method of generating LNOs is based on a low-rank approximation to the projection operator for the occupied space 
by a local eigendecomposition at each atomic site, and the band structure calculations with PAOs and LNOs clearly show 
that the resultant LNOs 
span well the occupied space of not only gapped systems but also metals. It is also worth mentioning that 
the computational cost of generating LNOs is almost negligible thank to the independent calculation at each atomic site. 
By replacing PAOs with LNOs in the long range region of the truncated cluster in the DC method, the computational cost of 
the DC method can be reduced without largely sacrificing the accuracy. 
Nothing that the DC-LNO method holds the simple algorithm of the original DC method suited to the parallel calculation, 
we have implemented a multilevel parallelization using MPI by taking account of the atom level, spin level, and 
diagonalization level. It was demonstrated that the speed-up of the DC-LNO method by the multilevel parallelization can 
be expected up to a specific number of MPI processes which corresponds to the product of the number of atoms, the multiplicity 
of spin index, and the number of CPU cores in a single computer node. For example, if a spin-polarized calculation is 
performed for a system consisting of 1000 atoms on a parallel computer with 20 CPU cores per node, a high parallel 
efficiency might be expected up to 40000 MPI processes. As a validation of the applicability of the DC-LNO method, 
we have performed MD simulations for liquid phases of an insulator, semi-conductor, and metals, and confirmed that 
the RDFs calculated by the DC-LNOs are in good agreement with those by the conventional $O(N^3)$ diagonalization method, 
which may lead to its various applications to structural determinations of amorphous and liquid structures of complicated 
materials \cite{Sakuda2017A,Sakuda2017B,Ileri2014,Koziol2017,Feng2018}.
Considering the simplicity and robustness of the algorithm, we conclude that the DC-LNO method is an efficient and accurate 
approach to large-scale DFT calculations for a wide variety of materials including metals. 

\begin{acknowledgments}
This work was supported by Priority Issue (Creation of new functional 
devices and high-performance materials to support next-generation 
industries) to be tackled by using Post 'K' Computer, MEXT, Japan.
Part of the computation in the study was performed using the computational
facility of the Japan Advanced Institute of Science and Technology.

\end{acknowledgments}

\appendix

\setcounter{figure}{0} \renewcommand{\thefigure}{A.\arabic{figure}}

\section{Asymptotic behaviors of the off-diagonal Green functions}

As an example we show asymptotic behaviors of the off-diagonal Green functions for an one-dimensional (1D) tight-binding (TB) 
model with a single $s$-orbital on each site, and relate the asymptotic behaviors to electronic structures
in gapped and metallic systems. The analysis interprets evidently the oscillating behavior of the error 
in the total energy of the metallic systems as shown in Fig.~5, and the rapid convergence 
in a high electronic temperature \cite{Ozaki2001,Krajewski2005}.

Let us consider an orthogonal chain model with the nearest neighbor interaction $t$ and the on-site energy $\varepsilon$ 
as defined by 
\begin{eqnarray}
  \hat{H} = \varepsilon\sum_{i}\hat{c}^{\dag}_i \hat{c}_i + t\sum_{i}\left( \hat{c}^{\dag}_i \hat{c}_{i+1} + {\rm h.c.} \right),
\end{eqnarray}
where $t$ is assumed to be positive. 
By tri-diagonalizing the Hamiltonian with a Lanczos algorithm starting from a site ($i=0$), and calculating 
the diagonal Green function via a continued fraction using the recursion method \cite{Haydock1975,Haydock1980}, 
one obtains a well known result for the diagonal Green function $G_{00}$ as follows:
\begin{eqnarray}
  G_{00}(Z) = \frac{1}{\sqrt{(Z-\varepsilon)^2-4t^2}}.
\end{eqnarray}
The off-diagonal Green functions can be obtained by using a recurrence relation \cite{Ozaki2000} derived from $G^{(L)}(Z)(Z-H^{(L)})=I$, 
where $G^{(L)}$ and $H^{(L)}$ are the Green function and Hamiltonian matrices represented by the Lanczos vectors, 
and by performing a back unitary transformation as 
\begin{eqnarray}
   \label{eq:G01}
  G_{01}(Z) &=& G_{00}(Z)\frac{\gamma}{2} - \frac{1}{2t},\\
   \label{eq:G02}
  G_{02}(Z) &=& G_{00}(Z)\left( \frac{\gamma^2}{2}-1 \right)-\frac{\gamma}{2t},\\
   \label{eq:G03}
  G_{03}(Z) &=& G_{00}(Z)\left( \frac{\gamma^3}{2}-\frac{3}{2}\gamma \right)-\frac{\gamma^2-1}{2t},\\
   \label{eq:G04}
  G_{04}(Z) &=& G_{00}(Z)\left( \frac{\gamma^4}{2}-2\gamma^2 + 1 \right)-\frac{\gamma^3-2\gamma}{2t},
\end{eqnarray}
where $\gamma=(Z-\varepsilon)/t$ and $G_{0j}$ is the off-diagonal element of Green function between the sites $0$ and $j$. 
It turns out that the off-diagonal Green functions can be expressed by $G_{00}$ and $\gamma$. To see asymptotic behaviors
of the off-diagonal Green functions, by employing the following formula \cite{Gradshteyn2007}:
\begin{eqnarray}
   \frac{1}{\sqrt{a^2-x^2}} = \frac{1}{a}\sum_{n=0}^{\infty} \frac{\binom{2n}{n}}{4^n a^{2n}}x^{2n}
   \label{eq:mathformula}
\end{eqnarray}
with the radius of convergence $\vert a \vert$, we Taylor exand $G_{00}$ at $\gamma^{-1}=0$ as 
\begin{eqnarray}
  G_{00}(Z)
      \nonumber
      &=& 
           \frac{1}{t} 
           \sum_{n=0}^{\infty}
           \frac{\binom{2n}{n}}{4^n 2^{-2n}}\gamma^{-(2n+1)},\\
      &=&
      \frac{1}{t} 
              \left(
                \frac{1}{\gamma} 
               +  
                \frac{2}{\gamma^3} 
               +  
                \frac{6}{\gamma^5} 
               +  
                \frac{20}{\gamma^7}
               + \cdots 
              \right), 
   \label{eq:G00expand}
\end{eqnarray}
where the convergence is guaranteed for $\vert \gamma \vert>2$.
By inserting Eq.~(\ref{eq:G00expand}) into Eqs.~(\ref{eq:G01})-(\ref{eq:G04}), and taking the leading terms, 
we obtain the following relation:
\begin{eqnarray}
  G_{0j}(Z) \propto \frac{1}{t \gamma^{j+1}}.  
   \label{eq:G0jexpand}
\end{eqnarray}
Thus, we see that $G_{0j}$ approaches to zero asymptotically for $\vert \gamma \vert>2$ as $j\to \infty$.
On the other hand the Green functions at $Z=\varepsilon$ corresponding to $\gamma=0$ are given by 
\begin{eqnarray}
  \label{eq:G00_at_a}
  G_{0(2k-1)}(\varepsilon) &=& \frac{(-1)^k}{2 t},\\ 
  G_{0(2k)}(\varepsilon) &=& (-1)^k G_{00}(\varepsilon),
\end{eqnarray}
where $G_{00}(\varepsilon)=\frac{-i}{2 t}$. It is found that $G_{0j}$ at $\gamma=0$ exhibits an oscillating 
behavior as a function of $j$, and never decays. 

\begin{figure}[t]
    \includegraphics[width=8.7cm]{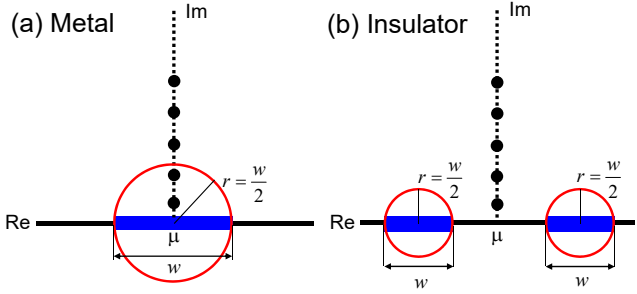}
    \caption{
     Relation between the position of Matsubara poles (black filled circels), the spectrum range 
     on the real axis (blue filled rectangles), and the convergent region of the off-diagonal Green functions whose 
     boundaries are shown by red circles in the complex plane for (a) metal and (b) insulator.
     In the simple TB model the band width $w$ is given by $4 t$, and the off-diagonal Green functions 
     $G_{0j}$ decays asymptotically as $j$ increases in the exterior region of the red circles.
     }
\end{figure}

We now relate the asymptotic behaviors of Green functions
to the calculation of density matrix which is defined by Eq.~(\ref{eq:dm}). Introducing the Matsubara expansion 
of the Fermi-Dirac function, and changing the integration path with the Cauchy theorem, one has \cite{Ozaki2007} 
\begin{eqnarray}
  \label{eq:rho0j}
  \rho_{0j} = \frac{1}{2}\delta_{0j}
             + {\rm Im}
             \left[ 
              \frac{2i}{\beta}\sum_{p=1}^{\infty} G_{0j}(\alpha_p)
             \right],
\end{eqnarray}
where $\alpha_p$ are Matsubara poles located at $\mu+i\frac{(2p-1)\pi}{\beta}$ 
with a chemical potential of $\mu$ and $\beta=\frac{1}{k_{\rm B}T}$. 
The expression allows us to figure out a relation between the Matsubara poles, where the Green functions are evaluated,
and the convergent region of the off-diagonal Green functions as illustrated in Fig.~A.1. 
Remembering that in the simple TB model the band width $w$ is given by $4 t$, and assuming that the single band
is half filled in the metallic case, we may have Matsubara poles in the red circle which is the non-convergent region 
of the off-diagonal Green functions as shown in Fig.~A.1(a). Since the off-diagonal Green functions evaluated at 
the Matsubara poles in the red circle do not simply decay in real space as $j$ increases, the truncation scheme commonly 
adopted in most of O($N$) methods should suffer from the long range characteristics of the off-diagonal Green functions, while 
the effect can appear in a different way depending on underlying principles of each O($N$) method. 
In the DC-LNO method the truncated eigenvalue problem $H^{(i)}c^{(i)}_{\mu}=\varepsilon^{(i)}_{\mu}S^{(i)}c^{(i)}_{\mu}$ is 
solved for each atom $i$, and the integrations of Eqs.~(\ref{eq:dm}) and (\ref{eq:edm}) can be easily performed on the real axis 
since we have the approximate spectrum representation of Eq.~(\ref{eq:Gi}). The way of evaluating the density matrix is numerically 
equivalent to the computational method via a generalized formula of Eq.~(\ref{eq:rho0j}) to the non-orthogonal basis set, 
where the Green functions for the truncated problem are computed at each Matsubara pole by the inverse calculation, 
since the Green function computed through the spectrum representation is exactly the same as one computed by the inverse calculation. 
Therefore, the oscillating behavior of error in the total energy calculation observed in Figs.~5(d), (e), and (f) should be 
attributed to the long range characteristics of the off-diagonal Green functions. 
It can also be understood that the use of a higher electronic temperature suppresses the deficiency since all the Matsubara poles can 
be placed in the exterior region of the red circles beyond a critical temperature \cite{Ozaki2001,Krajewski2005}. 
On the other hand we model an insulator by considering two bands as shown in Fig.~A.1(b), where each of them is expressed 
by the 1D TB model and the bands are separated by a finite gap. Unlike the metallic case, all the Matsubara poles are located 
in the exterior region of the red circles. The feature guarantees that $\rho_{0j}$ decays as $j$ increases since all the Green 
functions in the summation of Eq.~(\ref{eq:rho0j}) decay as $j$ increases, theoretically justifying that the truncation scheme 
is valid for gapped systems, although our benchmark calculations imply that the use of a large cutoff radius diminishes 
the effect of the long range characteristics of the off-diagonal Green functions even to metals at least for the calculations 
of density matrix and energy density matrix in a practical sense.


\begin{thebibliography}{99}
\bibitem{SVD}
To perform the low-rank approximation, a Schur decomposition which is a similarity transformation
can also be considered. However, an ordered Schur decomposition, which orders diagonal elements 
of the upper triangular matrix in descending order, is required to extract a subspace for a set 
of LNOs, and the algorithm has not been available in popular libraries such as LAPACK \cite{LAPACK}.
Another choice is a singular value decomposition (SVD), which has been widely used for low-rank 
approximations in many fields. While SVD might work in a similar way, it should be noted that  
SVD is not a similarity transformation, and the singular values do not correspond to the populations.
\end{thebibliography}
\end{document}